\documentclass{article}

\usepackage{arxiv}

\usepackage[utf8]{inputenc} 
\usepackage[T1]{fontenc}    
\usepackage{hyperref}       
\usepackage{url}            
\usepackage{booktabs}       
\usepackage{amsfonts}       
\usepackage{nicefrac}       
\usepackage{microtype}      
\usepackage{lipsum}		
\usepackage{multirow}
\usepackage{graphicx}
\usepackage[numbers,sort&compress]{natbib}
\usepackage{doi}
\usepackage{xurl}

\title{Identity Deepfake Threats to Biometric Authentication Systems: Public and Expert Perspectives}

\author{%
\href{https://orcid.org/0000-0003-3697-0706}{\includegraphics[scale=0.06]{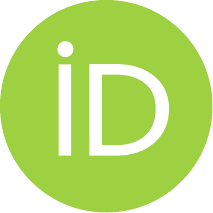}\hspace{1mm}Shijing He}\thanks{Both authors contributed equally to this research.} \\
King's College London\\
London, United Kingdom\\
\texttt{shijing.he@kcl.ac.uk} %
\And
\href{https://orcid.org/0000-0002-0697-7942}{\includegraphics[scale=0.06]{orcid.pdf}\hspace{1mm}Yaxiong Lei}\footnotemark[1]\thanks{Corresponding author.}\\
University of St Andrews\\
St Andrews, United Kingdom\\
\texttt{yl212@st-andrews.ac.uk} %
\And
\href{https://orcid.org/0000-0002-3129-3550}{\includegraphics[scale=0.06]{orcid.pdf}\hspace{1mm}Zihan Zhang}\\
University of St Andrews\\
St Andrews, United Kingdom\\
\texttt{zz66@st-andrews.ac.uk} %
\And
\href{https://orcid.org/0009-0000-8187-8939}{\includegraphics[scale=0.06]{orcid.pdf}\hspace{1mm}Yuzhou Sun}\\
University of St Andrews\\
St Andrews, United Kingdom\\
\texttt{ys95@st-andrews.ac.uk} %
\And
\href{https://orcid.org/0000-0001-5628-7328}{\includegraphics[scale=0.06]{orcid.pdf}\hspace{1mm}Shujun Li}\\
University of Kent\\
Canterbury, United Kingdom\\
\texttt{s.j.li@kent.ac.uk} %
\And
\href{https://orcid.org/0000-0003-3881-0546}{\includegraphics[scale=0.06]{orcid.pdf}\hspace{1mm}Chi Zhang}\\
University of St Andrews\\
St Andrews, United Kingdom\\
\texttt{cz38@st-andrews.ac.uk} %
\And
\href{https://orcid.org/0000-0002-2838-6836}{\includegraphics[scale=0.06]{orcid.pdf}\hspace{1mm}Juan Ye}\\
University of St Andrews\\
St Andrews, Fife KY16 9SX, United Kingdom\\
\texttt{jy31@st-andrews.ac.uk} %
}

\renewcommand{\shorttitle}{Identity Deepfake Threats to Biometric Authentication Systems}

\begin{document}
\maketitle

\begin{abstract}
	Generative AI (Gen-AI) deepfakes pose a rapidly evolving threat to biometric authentication, yet a significant gap exists between expert understanding of these risks and public perception. This disconnection creates critical vulnerabilities in systems trusted by millions. To bridge this gap, we conducted a comprehensive mixed-method study, surveying 408 professionals across key sectors and conducting in-depth interviews with 37 participants (25 experts, 12 general public [non-experts]). Our findings reveal a paradox: while the public increasingly relies on biometrics for convenience, experts express grave concerns about the spoofing of static modalities like face and voice recognition. We found significant demographic and sector-specific divides in awareness and trust, with finance professionals, for example, showing heightened skepticism. To systematically analyze these threats, we introduce a novel Deepfake Kill Chain model, adapted from Hutchins et al.'s cybersecurity frameworks to map the specific attack vectors used by malicious actors against biometric systems. Based on this model and our empirical findings, we propose a tri-layer mitigation framework that prioritizes dynamic biometric signals (e.g., eye movements), robust privacy-preserving data governance, and targeted educational initiatives. This work provides the first empirically grounded roadmap for defending against AI-generated identity threats by aligning technical safeguards with human-centered insights.
\end{abstract}


\twocolumn

\section{Introduction}
Recently, misinformation (deliberately deceptive or fabricated content designed to mislead and harm~\cite{aldwairi2018detecting,jang2018third,kumar2018false}) has undergone a significant transformation. At the center of these concerns is the rapidly advancing realm of deepfake technology, which employs generative AI (Gen-AI) to produce hyper-realistic synthetic content~\cite{pei2024deepfake,schmitt2024digital}. These forgeries range from fabricated web pages~\cite{deepfakecase2025} to synthetic human images nearly indistinguishable from reality~\cite{bray2023testing,lin2025omnihuman}, posing a fundamental challenge to digital trust. This trend is amplified by the popularity of video-centric platforms like YouTube~\cite{lee2021believe,pu2021deepfake} and image-centric portals such as Instagram and X/Twitter~\cite{bray2023testing,shahzad2022review}, and widespread use of video conferencing tools (e.g., Zoom)~\cite{kagan2023zooming}. The implications are far-reaching, threatening public discourse, individual rights, and broader societal trust~\cite{gregory2022deepfakes}.

Furthermore, biometric authentication has become increasingly adopted to verify identity (e.g., facial, or physiological authentication~\cite{sharma2022reliable, ometov2018multi}) but is paradoxically vulnerable to exploitation. While biometric systems are often proposed as robust defenses against identity forgery, their sensitivity makes them prime targets for deepfake attacks, thereby introducing significant privacy and ethical risks as stolen or replicated biometric data can enable malicious impersonation~\cite{rui2018survey}. This duality—where biometrics function as both a safeguard and a vulnerability—remains critically understudied. Prior research has broadly explored technical detection methods (e.g.,~\cite{guera2018deepfake,pan2020deepfake}), the cultivation of public awareness and critical thinking~\cite{bray2023testing, caldwell2020ai}, and deepfake risks in political misinformation and social manipulation~\cite{kugler2021deepfake, chapagain2024deepfake, meskys2020regulating, pawelec2024decent}. Yet, few studies have addressed how deepfakes specifically threaten biometric authentication systems, despite their widespread adoption~\cite{sharma2022reliable, ometov2018multi}. To address this gap, this paper investigates how both the \textbf{general public} and \textbf{experts} perceive the threats posed by Gen-AI identity deepfakes (image/video) to biometric authentication. We aim to answer the following research questions (RQs):

\begin{enumerate}
    \item[RQ1:] \emph{How do experts assess the effectiveness, limitations, and ethical considerations of biometric authentication solutions specifically designed to detect or mitigate biometric-related Gen-AI identity deepfake threats (image/video)?}
    \item[RQ2:] \emph{How does the general public perceive biometric-related Gen-AI identity deepfake threats (image/video), and to what extent do they trust or understand biometric authentication methods as protection against such impersonation attacks?}
\end{enumerate}

To investigate these RQs, we employed a mixed-method design~\cite{hong2018mixed}, combining a survey ($n$=408) with semi-structured interviews ($n$=37) with 25 expert participants (e.g., researchers and/or practitioners in biometric authentication, security \& privacy, or AI) and 12 general public participants (with no background in aforementioned fields). 

\textbf{Contribution.} Our study offers the first empirical investigation of expert and public perceptions regarding Gen-AI deepfake threats to biometric authentication systems. Whereas prior work has typically examined AI adoption in broad strokes, our research pinpoints sector-specific perceptions, revealing that formal education and professional longevity do not straightforwardly predict practical AI knowledge. Our cross-sector sample uncovers pronounced age, industry, and education gradients, non-linear relationships that smaller studies have missed, showing, for instance, that AI familiarity peaks among early career technologists yet declines in mid career managerial roles. 

Second, we extend Hutchins et al.’s threat model~\cite{hutchins2011intelligence} by introducing an expert-driven deepfake kill chain that identifies where Gen-AI compromises facial or voice recognition and exposes overlooked attack surfaces in multi-factor workflows. Building on this framework, we quantify public awareness of deepfake risks and systematically analyze privacy concerns and acceptance of using sensitive biometric data (e.g., facial expressions) for deepfake detection, highlighting the trade offs users accept for convenience and trust. 

Third, we advance a tri-layer (technical, social, and legal) mitigation framework that recommends dynamic, involuntary biometric signals (e.g., microsaccades and micro-expressions), editable gaze trajectories, on device storage with privacy preserving analytics, and harmonize regulatory mandates. Our empirical validation shows that such signals are markedly harder to spoof, yet they introduce practical usability constraints that designers must address. 

Finally, we propose a governance and education road map that pairs transparent consent interfaces with sector specific literacy programs, translating theoretical safeguards into deployable, ethically grounded, and privacy preserving biometric solutions.

\section{Related Work} \label{related_works}

\subsection{Deepfake Generation and Detection} \label{deepfake_detection}
Early deepfake methods primarily relied on convolutional neural networks (CNNs) for face/voice manipulation like face-swapping or lip synchronization~\cite{lyu2020deepfake, li2024identity}. These early iterations often exhibited detectable artifacts like unnatural blinking~\cite{jung2020deepvision}, mismatched head poses~\cite{becattini2024head}, or splicing edges~\cite{kumari2024image}, enabling identification through manual inspection. Advances in generative models, including Variational Autoencoders (VAEs)~\cite{kingma2013auto}, Generative Adversarial Networks (GANs)~\cite{karras2020analyzing, goodfellow2020generative} and diffusion models~\cite{blattmann2023stable, guo2023animatediff, liu2024towards, ma2024follow, lyu2024multimodal}, have since enabled highly realistic outputs, such as seamless facial composites and full-body avatars that are nearly indistinguishable from real footage~\cite{mink2024s,bray2023testing}. Innovations like free-form talking face generation~\cite{ma2023dreamtalk, zhou2019talking}, targeted facial attribute editing~\cite{huang2023ia, yang2021l2m}, and adversarial optimization or retraining~\cite{ninovic2024deepfake, pei2024deepfake} further enhance realism of deepfakes, complicating manual detection.

Spatial detection methods analyze frame-level inconsistencies, such as color distortions, pixel anomalies, or blending boundaries, where local inconsistencies or lighting mismatches may signal tampering~\cite{becattini2024head,kumari2024image,khamis2024perspectives}. Implicit identity cue comparisons with the expected facial structure for face-swapping detection~\cite{Huang_2023_CVPR}, multi-scale patch similarity modules to model fine-grained relationships among facial regions to capture subtle artifacts~\cite{Chen_Yao_Chen_Ding_Li_Ji_2021}, and attention-based frameworks to isolate anomalies~\cite{1zhuang2022uiavitunsupervisedinconsistencyawaremethod}. Multi-attentional architectures, representative forgery mining, and hybrid training approaches enhances the discriminative power of spatial detectors~\cite{Zhao_2021_CVPR,Wang_2021_CVPR,Wang_2023_CVPR}.

Temporal detection exploit the dynamic nature of video data to detect inconsistencies across frames. For instance, Thumbnail Layout (TALL) utilizes layout information from video thumbnails to capture structural cues that vary over time~\cite{10378447}. Zheng et al. proposed a fully temporal convolution network to measure inter frame consistency, flagging significant deviations as potential signs of manipulation~\cite{9710282}. Models that learn self-consistency explicitly enforce uniformity in feature representations over time so that disruptions serve as markers of deepfakes~\cite{9710718}. Further, approaches that jointly model spatial and temporal information provide robust detection by analyzing frame by frame anomalies that may be missed when only a single dimension is considered~\cite{gu2021spatiotemporal}.

Alternative strategies extends beyond traditional spatial or temporal analysis. For instance, LSDA uses latent space augmentation to overcome forgery specificity and improve generalization across diverse manipulation techniques~\cite{10657985}. Self-blending data augmentation and adversarial self-supervised frameworks enables models to learn more robust discriminative features~\cite{9880195,9879418}. Leveraging large-scale pretrained models like CLIP provides a powerful, transferable feature backbone for detection tasks~\cite{pmlr-v139-radford21a}; while space-time attention and masked autoencoders offers efficient means to capture complex video representations~\cite{NEURIPS2022_416f9cb3,bertasius2021space}. Frequency-domain methods detect GAN ``fingerprints'' from upsampling artifacts~\cite{zhang2019detecting,marra2019gans, yu2019attributing, durall2020watch}, while proactive watermarking or tags traces subsequent manipulations~\cite{asnani2022proactive, zhao2023proactive, wang2021faketagger, yu2021artificial}.

As conventional detection methods become increasingly vulnerable to more realistic and adaptive deepfake attacks, researchers are turning to biometric-based approaches that rely on intrinsic human characteristics~\cite{khan2025generative}, such as blinking patterns~\cite{jung2020deepvision}, inconsistent head poses~\cite{yang2019exposing}, micro-expressions~\cite{pei2024deepfake}, lip synchronization~\cite{haliassos2021lips, yang2020preventing}, and absent physiological cues~\cite{li2018ictu}. These physiological traits are deeply embedded in human physiology and are therefore difficult to fabricate with precision. Recent industry deployments illustrate the increasing reliance on biometric authentication with built-in defenses against deepfake threats. For instance, ID.me, widely adopted by U.S. government, combines facial recognition with liveness detection based on blinking and micro-expressions\footnote{\url{https://www.id.me/government}}. Worldcoin uses iris biometrics with on-device processing and cryptographic privacy guarantees to verify identity\footnote{\url{https://github.com/worldcoin/open-iris}}. These systems operate in adversarial settings, where deepfakes present real impersonation risks, and they rely on specific trust assumptions, such as secure client-side capture or hardware-based anchors.

\subsection{Public Perceptions of Biometric Data}
As biometric authentication (e.g., facial, behavioral recognition) becomes more integrated into various sectors, devices and systems, public attitudes reveal growing concerns over data sensitivity, surveillance, and loss of control over personal identity~\cite{prabhakar2003biometric}. These concerns have spurred support for stronger regulation, independent oversight, and strict limits on how biometric data is collected and used\footnote{\url{https://www.adalovelaceinstitute.org/report/beyond-face-value-public-attitudes-to-facial-recognition-technology/}}. For instance, Ritchie et al. found that while people generally support biometric use in criminal justice settings, such acceptance is conditional and highly context-dependent, with many expressing discomfort about widespread surveillance and a lack of transparency~\cite{ritchie2021public}. Similarly, Seng et al. revealed that user perceptions of facial recognition systems are nuanced, shaped by factors such as trust in the deploying entity, control over facial data, perceived utility, and the physical context of deployment~\cite{seng2021first}.

\subsection{Public and Expert Awareness of Deepfakes} \label{awareness_deepfake}

Deepfakes captured significant public and expert attention due to their potential to disrupt democratic processes, compromise public trust, and challenge the integrity of visual media~\cite{meskys2020regulating,battista2024political,kharvi2024understanding,kugler2021deepfake,wu2025understanding}. Recent studies highlighted significant variation in human abilities to detect artificially generated media across countries, underscoring the global and cross-cultural implications of deepfake threats~\cite{frank2024representative}. These artificial media were often perceived as authentic enough that viewers found AI-generated content more trustworthy than human-generated material, further complicating detection efforts~\cite{ha2024organic,nightingale2022ai}. Public awareness remained inadequate: many individuals were poorly informed about the risks associated with deepfakes. For instance, Caldwell et al. emphasized that deepfakes could enable identity theft, smear campaigns, election manipulation, and fraud, threatening public trust in digital content~\cite{caldwell2020ai}. Even simpler ``cheapfakes'' or ``shallowfakes'' demonstrated the capacity to sway public opinion through minimal editing~\cite{bray2023testing}. Chesney and Citron argued that deepfakes posed a critical challenge to privacy, democracy, and national security, urging immediate educational and policy interventions~\cite{chesney2019deep}. Wu et al. found that intelligence analysts, frustrated by fragmented single-purpose detectors, want a transparent end-to-end system that unifies analytics, safeguards data, and produces report-ready explanations~\cite{wu2025understanding}. 

Additionally, deepfakes complicated traditional forensic methods, prompting the need for advanced verification tools to authenticate video evidence in legal contexts~\cite{maras2019determining}. Expanding on privacy-preserving technologies, Khamis et al. found that deepfakes were perceived as equally effective as traditional obfuscation methods (e.g., blurring, pixelation) in privacy protection but integrated better aesthetically into photos, reducing visual disruption. However, deepfake requires ethical safeguards (e.g., marking manipulated content, synthetic faces) to prevent misuse~\cite{khamis2024perspectives}. Despite progress in detection techniques, technological advancements in content manipulation outpaced public awareness~\cite{westerlund2019emergence}. Researchers stressed the need to improve both technical detection and public understanding, noting that a well-informed citizenry could better resist manipulation through critical evaluation of digital media~\cite{farid2019confronting}.

\subsection{Positioning and Research Gap}
Prior work has tended to treat deepfakes chiefly as a misinformation threat and biometrics chiefly as a privacy issue; our study is the first to bridge these domains by combining a large cross-sector survey with expert interviews to map how Gen-AI deepfakes endanger biometric authentication in practice. We extend Hutchins et al.’s kill chain~\cite{hutchins2011intelligence} into a deepfake specific threat model; the resulting tri-layer mitigation and governance framework moves beyond detection accuracy to integrate privacy preserving storage, consent design, and targeted public education from technical, social, and legal perspectives.
\section{Methodology}
\label{method}
We conducted a mixed-method research~\cite{hong2018mixed} to achieve methodological triangulation and complementarity, including a survey study with key industry groups (e.g., finance, government) ($n$=408) and semi-structured interviews with 25 experts participants (EPs) and 12 general public participants (PPs) to investigate their awareness and perceptions regarding biometric authentication and Gen-AI identity deepfake threats. In this section, we introduce participants recruitment, study procedure, pilot study, and data analysis of our work. The ethics committee of the first author's university reviewed and approved this study.

\subsection{Recruitment}
We implemented the survey on Microsoft Forms and recruited participants from Prolific. Participants qualified for the user survey if they worked in the government sector, healthcare, the tech industry, finance, or academia, as these high-value sectors are more likely to be exploited for biometric authentication and security risks from Gen-AI identity deepfake attacks. This qualification check was conducted using a built-in filter provided by Prolific (e.g., working industries, experiences). We did not require prior experience with Gen-AI or biometric authentication. Each eligible participant also needed to: 1) be located in the UK; 2) aged 18 or over. The average completion time for the survey was 9 mins, and each valid response was compensated with £1.5. After removing low-quality responses, such as completed the survey significantly faster than average were removed (under 5 mins, compared to the average completion time), 408 responses were considered valid (see Table \ref{tab:Survey_demographic}).

We attracted 12 PPs by posting a recruitment ad on the first co-author's university’s mailing list. All potential participants filled out a screening survey. As we progressed through the interviews, few new insights emerged after the seventh interview. We continued with five more and observed nothing new to reach saturation~\cite{guest2006many}. 2 were interviewed in Mandarin, 10 were conducted in English. All interviews lasted 49 mins on average (see Table \ref{tab:PP_demographic}).

For EP interviewees, we aimed to secure a mix of researchers and practitioners. Eligibility criteria included: 1) being aged 18 or older with peer-reviewed publications in relevant computer science areas (biometric authentication, security \& privacy, or AI); 2) having practical experience in designing, implementing, or evaluating biometric authentication or Gen-AI systems. There were no restrictions based on gender, affiliation, or location. Initially, we searched relevant publications on privacy \& security issues involving Gen-AI deepfake threats to biometric authentication systems using Google Scholar, ACM Digital Library, and IEEE Xplore, generating a list of 32 potential interviewees from the UK. Our initial email outreach to 32 authors yielded a low response rate (9.4\%). To achieve our target sample size, we supplemented this with a snowball recruitment strategy~\cite{parker2019snowball}, which leveraged professional networks and social media to recruit an additional 22 experts from around the world. After 25 interviews, we reached theoretical saturation~\cite{guest2006many} and ceased recruitment. Among these, 19 interviews were conducted in English and 6 in Mandarin, with an average duration of 51.4 mins (see Table \ref{tab:EP_demographic}). Further, we compensated all EP and PP interviewees with a £10 gift card, in line with the UK's minimum wage standards and the interviewees' time commitment.

Our public survey and interviews exclusively involved UK-based participants to deeply examine perceptions within a specific regulatory and societal context (e.g., the GDPR and the UK Data Protection Act 2018). We selected industries (finance, healthcare, government, academia, and tech) given their critical reliance on biometric authentication and susceptibility to targeted Gen-AI identity deepfake threats. The UK-based public sample was chosen due to its stringent data protection frameworks, while EPs were recruited globally to capture broad technical insights and international best practices concerning biometric authentication and Gen-AI deepfake threats. Although this introduces geographic variation between participant groups, it strengthens the study by combining context-specific public views with globally relevant expert perspectives. 

\begin{table*}[ht]
    \centering
    \fontsize{6pt}{6pt}\selectfont
    \resizebox{0.8\textwidth}{!}{%
    \begin{tabular}{lllllll}
    \toprule
    \textbf{Age} & \textbf{Gender} & \textbf{Education} & \textbf{Industry} & \textbf{Work Experience} & \textbf{Gen-AI Experience (times)} & \textbf{Biometric Authentication Experience (times)} 
    \\
    \midrule
    8.8\% 18-24         & 57.1\% Male     & 7.1\% Doctoral        & 12.3\% Academia & 4.9\% < 1 yr & 15.9\% Never & 3.2\% Never 
    \\
    43.9\% 25-34          & 40.9\% Female   & 22.8\% Master    & 31.1\% Gov./Public      & 17.4\% 1--3 yrs & 25\% Occasionally (1--3/month) & 11.5\% (1--3/month) 
    \\
    25\% 35-44        & 2\% Non-binary & 65.7\% Bachelor  & 26\% Tech         & 16.7\% 3--5 yrs        &  16.9\% Regularly (1--3/week)  & 5.1\% Regularly (1--3/week) 
    \\
    13\% 45-54        &                & 4.4\% < High School       & 12\% Healthcare   & 24.8\% 5--10 yrs & 21.8\% Daily use (1–3/day) & 29.2\% Daily use (1–3/day) 
    \\
    9.3\% 55+          &                &                             & 18.4\% Finance            & 36.2\% 10 yrs+              & 20.4\% Multiple/day (3+/day) & 51\% Multiple/day (3+/day) 
    \\
    \bottomrule
    \end{tabular}}
    \caption{Survey respondents demographics.}
    \label{tab:Survey_demographic}
\end{table*}

\renewcommand{\arraystretch}{1}
\begin{table*}[!ht]
    \centering
    \fontsize{6pt}{6pt}\selectfont
    \resizebox{0.8\textwidth}{!}{%
    \begin{tabular}{llllllll}
    \toprule
        ~ & \textbf{Age} & \textbf{Gender} & \textbf{Education} & \textbf{Ethnic group} & \textbf{Background} & \textbf{Biometric Authentication Experience} & \textbf{Gen-AI Experience} \\ 
    \midrule
        PP1 & 35-44 & Male & Bachelor & Asian & Linguistic study & Face, Fingerprint & No experience\\
    
        PP2 & 18-24 & Female & Bachelor & Asian & Heritage and architecture & Face, Fingerprint, Voice & Text, Image, Audio, Video\\
    
        PP3 & 25-34 & Female & Master & Asian & Film studies & Face, Fingerprint, Voice, Iris & Text, Image, Video\\
      
        PP4 & 35-44 & Female & Doctoral & Asian & Chemistry & Face, Fingerprint, Voice & No experience\\
     
        PP5 & 25-34 & Female & Master & Black & Digital Business & Face, Fingerprint & Text \\
     
        PP6 & 18-24 & Male & Bachelor & White & Media management & Face, Fingerprint, Vein & Text, Image\\
     
        PP7 & 25-34 & Male & Bachelor & Asian & Linguistic study & Face, Fingerprint, Iris & Text\\
     
        PP8 & 18-24 & Female & Master & White & Economics and management & Face, Fingerprint & Text, Audio\\
     
        PP9 & 18-24 & Male & Bachelor & White & Physics & Face, Fingerprint, Voice & Text, Image, Video\\

        PP10 & 18-24 & Male & Bachelor & Black & Geography & Face, Fingerprint, Voice & Text, Image, Audio, Video \\

        PP11 & 35-44 & Male & Master & White & Sociology & Face, Fingerprint & Text, Image \\

        PP12 & 25-34 & Female & Bachelor & White & Media & Face, Fingerprint, Voice & Text, Image, Audio, Video \\
    \bottomrule 
    \end{tabular} }
    \caption{PP demographics, including their educational background, experience with biometric authentication and Gen-AI.}
    \label{tab:PP_demographic}
\end{table*}

\renewcommand{\arraystretch}{1}
\begin{table*}[!ht]
    \centering
    \fontsize{6pt}{6pt}\selectfont
    \resizebox{0.8\textwidth}{!}{%
    \begin{tabular}{llllllll}
    \toprule
        ~ & \textbf{Age} & \textbf{Gender} & \textbf{Education} & \textbf{Ethnic} & \textbf{Role} & \textbf{Institution} & \textbf{Academic Background} \\ 
    \midrule
        EP1 & 35-44 & Male & Doctoral & Asian & Associate professor & UK university & AI, Security \& Privacy \\
    
        EP2 & 25-34 & Female & Doctoral & White & Postdoc & US university & Security \& Privacy \\
    
        EP3 & 25-34 & Male & Doctoral & White & Assistant professor & EU university & Biometric Authentication, Security \& Privacy \\
      
        EP4 & 25-34 & Female & Doctoral & Asian & Postdoc & UK university & AI, Security \& Privacy\\
     
        EP5 & 25-34 & Male & Doctoral & Asian & Postdoc & UK university & AI\\
     
        EP6 & 25-34 & Male & Doctoral & White & Postdoc & EU university & Biometric Authentication, Security \& Privacy \\
     
        EP7 & 25-34 & Female & Doctoral & Black & Research scientist & EU university & Biometric Authentication, Security \& Privacy \\
     
        EP8 & 35-44 & Male & Doctoral & Asian & Assistant professor & US university & Biometric Authentication\\
     
        EP9 & 35-44 & Female & Doctoral & Asian & Associate professor & UK university & Security \& Privacy \\

        EP10 & 25-34 & Male & Doctoral & White & Research scientist & UK university & AI, Security \& Privacy \\
     
        EP11 & 25-34 & Male & Doctoral & Asian & Postdoc & CN university & AI\\
     
        EP12 & 18-24 & Female & Master & Asian & PhD candidate & CN university & Security \& Privacy\\
     
        EP13 & 18-24 & Female & Master & Asian & PhD candidate & CN university & AI, Security \& Privacy\\
     
        EP14 & 25-34 & Male & Doctoral & Asian & Assistant professor & CN university & AI, Biometric Authentication\\
    
        EP15 & 35-44 & Male & Doctoral & Asian & Full professor & CN university & Biometric Authentication, Security \& Privacy\\
     
        EP16 & 25-34 & Female & Doctoral & Asian & Assistant professor  & CA university & Biometric Authentication \\
             
        EP17 & 25-34 & Female & Doctoral & White & Postdoc & US university & AI, Security \& Privacy \\
             
        EP18 & 25-34 & Male & Doctoral & Asian & Research scientist & UK university & Biometric Authentication \\

        EP19 & 35-44 & Male & Doctoral & White & Postdoc & US university & Biometric Authentication, Security \& Privacy\\

        EP20 & 35-44 & Female & Doctoral & Asian & Associate professor & US university & Biometric Authentication, Security \& Privacy \\

        EP21 & 25-34 & Male & Doctoral & White & Assistant professor & US university & Biometric Authentication\\

        EP22 & 25-34 & Male & Doctoral & White & Postdoc & US university & Biometric Authentication, Security \& Privacy\\

        EP23 & 35-44 & Female & Doctoral & Black & Assistant professor & UK university & AI, Security \& Privacy\\

        EP24 & 35-44 & Male & Doctoral & White & Postdoc & EU university & Biometric Authentication\\

        EP25 & 25-34 & Male & Master & Asian & PhD candidate & EU university & AI, Biometric Authentication \\
    \bottomrule 
    \end{tabular}}
    \caption{EP demographics, including their roles, institutions, and academic backgrounds.}
    \label{tab:EP_demographic}
\end{table*}

\subsection{Survey Design} \label{survey_design}

To investigate how Gen-AI identity deepfake threats intersect with biometric authentication security, we designed a structured questionnaire aligning closely with our RQ2. The survey began with a consent form outlining the purpose of the study, followed by demographic questions capturing age, gender, education, work experience, and employment sector (e.g., tech, finance, academia). The questionnaire consisted of five primary factors:

\textbf{F1: AI Familiarity} focused on participants' familiarity with Gen-AI tools (e.g., ChatGPT, Stable Diffusion, GANs) and their perceived realism of deepfake-generated content. 

\textbf{F2: Enterprise/Industry AI Readiness} examined participants' perspectives on Gen-AI's dual role in cybersecurity, both for enhancing defenses and facilitating sophisticated attacks especially identity deepfakes (image/video). Respondents assessed their industry's readiness, effectiveness of Gen-AI-based identity deepfake threat detection systems, and the impact of regulatory frameworks. 

\textbf{F3: Trust in Biometrics} explored the frequency and contexts of biometric authentication use (e.g., smartphones, government services), preferred biometric methods, and overall trust levels. 

\textbf{F4: Confidence in Biometric Security in Gen-AI Deepfake Threats} specifically assessed participants' perceptions of biometric vulnerabilities against Gen-AI deepfake impersonation attacks. Items measured their confidence in current biometric authentication systems' security and the expected effectiveness of future biometric advancements integrated with Gen-AI.

\textbf{F5: Ethical AI Adoption} examined broader views on ethical Gen-AI practices, and the significance of continuous training from public sectors and companies, as well as views on international regulatory collaboration.

\subsection{Interview Procedure}
The first three co-authors (all native Mandarin speakers) conducted interviews remotely via Microsoft Teams. We obtained signed consent forms from participants and recorded all sessions with their consent. The recordings were transcribed using Notta.ai, and each transcript was independently reviewed by the same three authors.

\subsubsection{Interview Procedure for EPs} 
Each interview began with an overview of the study’s aims and warm-up questions, encouraging the interviewee to share research experiences and prior projects in biometric, deepfake, or Gen-AI development. We explored trade-offs between hardware authentication (e.g., hardware keys) and biometric authentication approaches (e.g., facial, fingerprint) in terms of security \& privacy, and usability. We also examined how organizations and governments store biometric data (e.g., cloud storage, user consent, and data protection) in sensitive contexts like banking or government services, to probed real-life examples of compromised systems or social engineering attacks using Gen-AI deepfake on personal identity. Further, we showed the human body display diagram to systematically examine currently available biometric data types, thereby developing a framework for analyzing the falsifiability of biometric data. In terms of human body structure, biometrics can be classified according to head, torso, and limbs. Head-based biometrics are the most numerous, comprising facial, iris, pupil, voice, and mouth shape recognition; extremity-based biometrics include palm print and fingerprint recognition; whole-body biometrics incorporate gait recognition. We also explored their views on AI-generated data (e.g., synthetic images or voices), focusing on potential manipulation of biometric signals. Ethical, legal, and global regulatory frameworks (e.g., the GDPR, the EU AI Act) were also addressed, noting how these regulations influence AI deployment. Lastly, we showed selected Gen-AI deepfake images~\cite{bray2023testing} and videos~\cite{xu2024vasa}, asking them to differentiate real vs. synthetic visuals and discuss cues for detecting fakes. We also examined how future biometric authentication and Gen-AI deepfake could address emerging security and privacy challenges.

\subsubsection{Interview Procedure for PPs} Similar to the EP interviews, we began by introducing the study’s aims and assessing participants' familiarity with biometric technologies and Gen-AI. We then explored the interviewees' authentication preferences and opinions on data collection, cloud storage, and potential misuse by companies or government entities. Additionally, we asked about their views on Gen-AI deepfake images and videos and the risks they pose to biometric systems. Finally, we presented the same Gen-AI deepfake images and videos used in the EP interviews~\cite{bray2023testing,xu2024vasa}, discussing the cues participants used to identify fakes and whether they believed current Gen-AI deepfake could convincingly deceive biometric authentication methods.

\subsection{Pilot Study}
Before launching the main survey, we ran two rounds of pilots (one round with friends and families, another rounds with actual 20 participants on Prolific) to collect participants’ feedback on potential improvement on the survey design.  
This pilot data was not included in our final results. Further, we ran a pilot study with two expert interviews (one postdoc in security \& privacy, and one research fellow in AI, both had biometric authentication research experience) and three public participant pilots (two from business/marketing, and one from gender studies) to ensure the questions were understandable and to identify any potential issues in the interview guide before proceeding with the main study. These five pilots were not included in the final analysis. 

\subsection{Data Analysis} 

Our surveys primarily consisted of 5-point Likert scales. After data collection, the first three co-authors thoroughly reviewed the dataset multiple times to become familiar with the responses and simultaneously filter out any low quality data. See detailed analysis in \S \ref{survey_results}. For qualitative interview data, we adopted an inductive thematic analysis approach~\cite{braun2006using} to examine all transcripts. First, to familiarize themselves with the data, the first three co-authors each closely read and independently coded the same two transcripts (selected at random, one EP, the other PP). During this initial coding phase, each of the three authors created a separate codebook. Next, they met to review and reconcile any discrepancies, merging similar codes, removing overlaps, and jointly agreeing on a codebook. They then tested the merged codebook by independently coding an additional transcript (different from the first two) for EPs and PPs separately, and reconvened to discuss and refine codes based on any new conflicts or ambiguities. After repeating this process two more times, the three authors reached code saturation, finding no further need to modify the codebook. Once the final codebook was established, the first author coded all EP \& PP transcripts, while the second author coded all EP transcripts, and the third author coded all PP transcripts. Finally, the team reviewed and organized the collective codes into themes and sub-themes relevant to the study’s RQs, carefully examining participant excerpts to define and refine how each theme was represented.

\section{Findings}
\label{findings}
In this section, we present the key findings in our survey analysis (see \S \ref{survey_results}), and key themes we observed across our qualitative interviews with EPs and PPs (see \S \ref{interview_results}). 

\subsection{Survey Results} \label{survey_results}
\begin{figure*}[ht]
  \centering
  \includegraphics[width=0.78\linewidth]{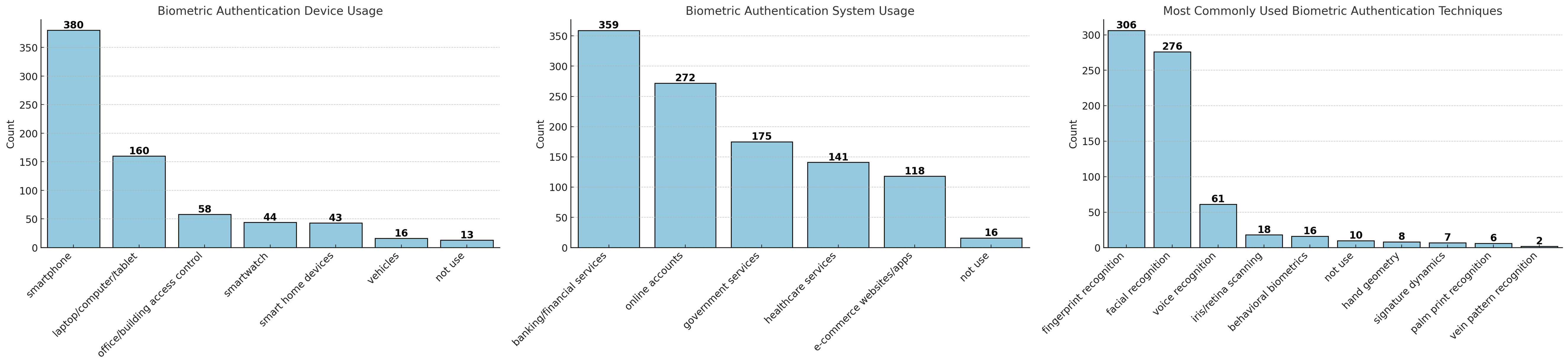} 
  \caption{Distribution of biometric authentication usage across devices, systems, and techniques.}
  \label{fig:gen_bio_exp}
\end{figure*}

Among these 408 valid responses, 51\% reported using biometric authentication multiple times a day; 20.4\% reported using Gen-AI multiple times a day. 93.1\% of respondents indicated that the most frequently used device for authentication was a smartphone, and 88\% of respondents used biometric authentication for banking and financial services (e.g., banking apps). Regarding biometric experience as shown in Figure \ref{fig:gen_bio_exp}, 75\% of participants had used fingerprint authentication, and 67.6\% had used facial authentication, while only 0.5\% had used vein authentication.

We performed one-way ANOVA to examine the relationship between demographic variables and our five factors. Only F1 demonstrates significant heterogeneity across nearly all demographic variables except gender, and also exhibits a clear generational gradient: younger respondents report substantially higher familiarity than older cohorts (progressively decreasing with age, particularly after age 35), reflecting more recently educated professionals entering the workforce with greater exposure to Gen-AI identity deepfake threats on biometric authentications, where technological familiarity may diminish with career progression into management or specialized roles (see Table \ref{tab:significance_summary}).

\begin{table}[ht]
\centering
\resizebox{0.48\textwidth}{!}{%
\fontsize{6pt}{6pt}\selectfont
\begin{tabular}{lcccccc}
\toprule
\multirow{2}{*}{\textbf{Demographic}} & \multicolumn{5}{c}{\textbf{Factor}} & \multirow{2}{*}{\textbf{Total}} \\ 
\cmidrule(lr){2-6} 
 & \textbf{F1} & \textbf{F2} & \textbf{F3} & \textbf{F4} & \textbf{F5} & \\ 
\midrule
Age & *** &  &  &  &  & 1 \\ 
Gender &  &  &  &  &  & 0 \\ 
Education Levels & *** &  & 0.002** &  & 0.005** & 3 \\ 
Working Industry & *** & 0.021* &  & 0.007** & *** & 4 \\ 
Working Experience & *** &  &  &  &  & 1 \\ 
\midrule
\textbf{Total} & 4 & 1 & 1 & 1 & 2 & \\ 
\bottomrule
\multicolumn{7}{p{0.95\linewidth}}{We used one-way ANOVA to test differences across demographic groups. Significance codes: *** p $<$ 0.001, ** p $<$ 0.01, * p $<$ 0.05. Exact p-values are shown where 0.001 $<$ p $<$ 0.05.} \\
\end{tabular}}
\caption{Summary of significant demographic differences}
\label{tab:significance_summary}
\end{table}

Working industry emerged as the most influential demographic factor, showing significant associations with F1, F2, F4, and F5 (95\%-99.9\%): 1) Academia and Tech Industry respondents reported significantly higher in F1 compared to Finance Industry, Government/Public Sector, and Healthcare. However, no significant difference was observed between Academia and Tech Industry, while Government/Public Sector demonstrating the lowest familiarity; 2) Only Healthcare professionals reported significantly higher perceptions of industry readiness (F2) compared to those in Academia; 3) Academia reported significantly lower confidence in biometric security and Gen-AI identity deepfake threats towards to biometric authentications, compared to Finance Industry and Government/Public Sector (see Table \ref{tab:posthoc_f1_age}) (F4); and 4) Academia reported significantly lower ethical perceptions in F5 compared to all other sectors (see Table \ref{tab:posthoc_f5_workingarea}). We found no significant differences among non-academic sectors, suggesting that academic professionals maintain more critical evaluations of current ethical AI practices.

Education levels significantly impact F1, F3, and F5: we found that doctoral degree holders reported significantly higher trust in biometrics (F3) compared to other education levels, while no significant differences were observed among other education groups; bachelor's degree holders reported significantly higher AI familiarity compared to doctoral and master's degree holders (see Table \ref{tab:posthoc_f1_education}); while bachelor's degree holders reported significantly lower ethical perceptions in F5 compared to doctoral degree holders. 

Meanwhile, working experience significantly influenced only F1 (with marginal effects on F4), indicating that mere professional longevity does not automatically translate to divergent perspectives on emerging technologies without specific exposure or training. Professional sector remains a critical determinant of AI perceptions, with Academia and Tech Industry demonstrating higher AI familiarity (F1) yet Academia showing more critical perspectives on F5 and F4, highlighting how academic environments may foster more rigorous evaluation of emerging technologies. 

Our results also revealed strong interconnections (see Tables \ref{tab:regression_results}, \ref{tab:mediation_results}, and \ref{tab:interaction_results}). Technology familiarity and acceptance, and industry AI readiness were closely linked, strongly influencing each other bidirectionally. In contrast, trust in biometrics showed weaker but meaningful connections, primarily influencing respondents' confidence in biometric security through perceptions of industry readiness. Familiarity significantly mediated the relationships between industry readiness and both confidence and ethical perceptions. Industry readiness mediated the relationship between trust and confidence, highlighting that trust shapes respondents' confidence mainly through industry preparedness. Confidence in biometric security also significantly affected how trust levels in biometrics related to ethical perceptions.

Although trust and ethical perceptions independently enhanced industry readiness, their combined effect was less than expected. Similarly, trust and readiness independently boosted ethical perceptions, yet jointly provided reduced benefits. Higher trust levels could even limit additional confidence gains when combined with AI familiarity. While ethical perceptions and industry readiness separately could reduce respondents' acceptance and familiarity, together they improved acceptance, indicating that ethical practices might offset readiness concerns. Finally, greater confidence amplified the positive influence of industry readiness on acceptance and familiarity.

\subsection{Interview Results} \label{interview_results}
\subsubsection{Deepfake Heightens the Risk of Biometric Authentication} \label{deepfake_threats}
All of our EPs first highlighted that the barrier to creating deepfakes has fallen dramatically. What once required weeks of model training can now be done in minutes on cloud services or even smartphones. Many customer free apps (e.g. DeepFaceLive\footnote{\url{https://www.deepfakevfx.com/downloads/deepfacelive/}}, Wav2Lip\footnote{\url{https://www.wav2lip.org/}}) enable real-time face swapping or voice morphing during video calls. These technical advances imply that deepfakes can reliably impersonate individuals in image, video, and voice media, easily bypassing casual human scrutiny and basic authentication systems. The result is a potent capability for attackers to fabricate almost any scenario on demand. Some EPs then outlined a clear progression in deepfake image and video development: 1) pixel-level consistency, primarily addressing static visual inconsistencies and background fragmentation; 2) human domain unification, where current technologies improve consistency at the video level using diffusion and flow match models to bridge gaps between human and forged action domains; 3) action domain approximation, involving the approximation of human actions for specific tasks; and 4) individualized action domain approximation, the final stage targeting individual-specific action patterns after mastering general human movements. 

\textbf{Escalating risks of deepfake attacks.} Nearly all EPs agreed that the advancement of Gen-AI deepfake images and videos has significantly amplified the risks associated with biometric authentication fraud. Also, the consensus among most EPs pointed towards a critical need for standardized benchmarks and more comprehensive training datasets to effectively evaluate and enhance the robustness and reliability of deepfake detection systems. Some cited the notable deepfake fraud incident in Hong Kong as clear evidence of the financial and reputational damage resulting from compromised biometric systems and human perception\footnote{British engineering firm Arup lost £20 million after a Hong Kong employee was deceived by deepfake video call: \url{https://edition.cnn.com/2024/05/16/tech/arup-deepfake-scam-loss-hong-kong-intl-hnk}}. They noted that adversaries first harvest public multimedia traces of high-value personnel (e.g., board members, system operators) from social media, webinars, and investor calls. Even sub-minute voice snippets or a handful of profile pictures suffice to train state-of-the-art diffusion or NeRF models; attackers fine-tune generative models to produce identity-faithful voice, image, or video segments. EP19 stressed that inexpensive cloud GPUs reduce training time \textit{``from days to hours''}, enabling rapid iteration until a human evaluator cannot reliably discern artifacts. 

In addition, most of our EPs agreed that deepfakes pose potentially serious threats across critical infrastructure sectors. For instance, EP7 and EP12 noted that some manipulated videos impersonating officials could trigger shutdowns or contaminate water supplies, leading to public health crises and grid instability. In the case of transportation, several EPs also pointed out that the healthcare sector and telecom systems are vulnerable to fraudulent announcements, which can spark panic and spread misinformation. Some EPs further emphasized that evolving deepfake technologies are enabling increasingly realistic synthetic media, which challenge the effectiveness of traditional biometric authentication methods. For example, EP22 highlighted the specific challenge posed by adversarial attacks: \textit{``Attackers don't even need comprehensive knowledge about the targeted systems. Simple manipulations such as compression or noise addition, or called laundering attacks [...] these can degrade the performance of forensic detection methods.''} 

Several EPs noted concerns about the limited robustness of current forensic detection techniques. As EP23 stated, \textit{``The forensic community largely assumes benign conditions, neglecting potential adversarial actions designed explicitly to mislead forensic analysis.''} They noted some deep learning-based detection methods faces inherent vulnerabilities due to lack of generalization beyond training environments, making them susceptible to various adversarial manipulations. Moreover, a few EPs emphasized the computational and resource challenges, as EP4 citing the International AI Safety Report 2025\footnote{\url{https://www.gov.uk/government/publications/international-ai-safety-report-2025/international-ai-safety-report-2025}} and noting, \textit{``Real-time detection of high-resolution, complex deepfake videos is computationally demanding, posing substantial barriers to effective implementation in practical applications.''} 

A few EPs noted that single-frame face unlock and one-shot voice prints were considered trivially spoofable once high-resolution samples are available. EP20 specifically described certain actors with limited resources who leverage public SaaS platforms (e.g., D-ID\footnote{\url{https://www.d-id.com/}}) for deepfake video generation. This highlights that authentication systems using single-factor verification may be vulnerable to such attacks, or even exploited in romance scams (e.g.,~\cite{wang2024victim}). EP20 added that some professional criminal groups, or even state-sponsored actors, are integrating deepfakes with disinformation campaigns and Operational Technology (OT) intrusions, thereby \textit{``weaponizing synthetic videos'' (e.g.,~\cite{pantserev2020malicious})}.

\textbf{Detection challenges in deepfake faces.} Most EPs emphasized that biometric modalities relying on static or easily replicable signals, such as facial images or voice samples, face significant vulnerabilities. Nearly one-third of EPs indicated that techniques like GANs, VAEs, and diffusion models have enabled the creation of synthetic replicas with remarkably realistic qualities. Additionally, EPs discussed the potential for sophisticated deepfake technologies to mimic subtle signals used in liveness detection, such as eye blinks and micro-expressions, making biometric verification increasingly challenging. 
Experts warned that this reliance on physiological cues is becoming a liability. As EP9 explained, the assumption that AI cannot replicate such subtleties is now outdated: \textit{``Deepfake detection often depends on spotting tiny physiological cues that are thought to be hard for AI to fake. While that might work [sic] sometimes, today’s advanced deepfakes can already mimic things like micro-expressions and subtle eye movements pretty convincingly, making these detection methods a lot less reliable.''} In particular, several EPs warned that, once a deepfake bypasses an initial authentication access, attackers move quickly to credential dumping, privilege escalation, and lateral movement, often targeting industrial-control or other high-value segments. As EP8 stated, \textit{``the deepfake is usually the only bespoke component […] everything after that is just commodity malware.''} EP23 added that, to preserve persistence and hamper forensic analysis, adversaries frequently recycle the same cloned persona in later sessions or launder the media (e.g., through compression or resizing).

Most EPs pointed out the rapid technological evolution of deepfake generation, which continuously produces increasingly realistic and difficult-to-detect videos. For instance, nearly one third highlighted that any single detection method struggles to remain effective across all scenarios. Variations in resolution, compression levels, and sharing platforms further complicate consistent detection. EP24 also addressed the robustness issue inherent in deep learning-based detection systems, stating, \textit{``Deep learning models are the core of most current detection methods but often suffer robustness problems. They works well in training environments but in real-world scenarios due to adversarial attacks and the intrinsic variability of deepfake video quality.''}

Some EPs discussed the absence of comprehensive benchmark datasets severely restricts our ability to objectively evaluate and enhance detection methodologies, much work remains to establish standardized testing conditions. They also discussed clues for identifying deepfake images, noting unnatural background transitions, misaligned facial muscle movements, and discrepancies in lighting reflections within eyes. EP17 specifically stated, \textit{``Eyes are often revealing—genuine human eyes reflect natural lighting consistently. In contrast, deepfake-generated eyes frequently appear unnaturally flat or overly smooth, disrupting the expected natural reflections.''} Several EPs noted that current detection approaches are computationally intensive and require extensive resources, impeding real-time applications. They advocated the development of data efficient models to mitigate the heavy computational demands and large training datasets traditionally required by deep learning approaches.

\textbf{Static and time-series biometrics.}
Nearly all EPs mentioned that current Gen-AI deepfake images mainly focuses on face swapping, or by replacing the face and mimicking the voice for login and social engineering attacks. Some noted the ubiquity of security cameras creates significant risk of head and body information exposure. Facial information represents the most immediately vulnerable biometric due to its accessibility and is consequently the primary target for deepfake replacement. While limb behavior and walking posture can also be compromised, these modalities require longer data collection periods to extract the involuntary micro-features that characterize authentic human movement. Similarly, as deepfake completes the challenge of inconsistent pixel levels in the face and the environment, there will be less and less information available to detect and recognize forgeries, such as using information from the eyes even though it requires specialized equipment.

Some EPs highlighted the critical distinction between static and dynamic biometric data. Facial recognition based on still photographs is substantially more vulnerable to deepfake than authentication systems utilizing facial micro-expressions captured in video sequences. For instance, EP1 noted that future authentication systems and deepfake detection mechanisms should prioritize time-series data (dynamic) over static data, stating that: \textit{``So thinking along these lines, future authentication systems, or in-depth forgery monitoring tools, could use more time-series data, which is dynamic rather than static data.''} They also summarized these promising dynamic biometric modalities, including gait analysis (walking patterns), limb movement characteristics, non-conscious micro-movements during computer mouse operation, head movement behavioral patterns, facial micro-expression sequences, and eye movement trajectories. These time-series biometrics capture unconscious behavioral patterns that are significantly more difficult to forge than static biometric snapshots, potentially offering more robust security against increasingly sophisticated deepfake attacks.

\textbf{Involuntary biometrics against identity deepfake threats.} Some EPs discussed the potential of micro-face expressions and eye-movement-based methods. While advanced Gen-AI deepfake can now imitate faces or voices, micro face expressions and eye-based signals (e.g., micro-saccades) remain relatively difficult for deepfake technology to replicate. As EP7 noted: \textit{``I think eye movement patterns are better than micro-face expressions and suited for high-security applications, like accessing classified data or sensitive facilities.''} EP1 noted the unique security advantage of eye movements, stating, \textit{``Unlike facial authentication, where users can be tricked into altering their actions, while micro eye movements cannot be consciously controlled in response to external stimuli. This makes replay attacks ineffective, as deepfake now cannot replicate or manipulate reflexive eye responses. Also, previously captured eye movement data cannot be reused.''} 

Additionally, a few EPs specifically highlighted the usability of involuntary biometrics, emphasizing that a ``gaze'' approach requires minimal user effort and could be particularly beneficial for individuals with limited mobility or those who frequently interact with devices. Nonetheless, they highlighted practicality challenges in authentication time spent, limiting widespread adoption in consumer devices. Like EP1 stated, \textit{``If this gaze data takes minutes to authenticate, it’s not realistic for everyday use.''} EP1 further discussed the difficulty of gaze data collection, noting: \textit{``Implicit data, by its nature, isn’t easily accessible or observable. For example, facial data is quite difficult to hide, many people don’t cover their faces all the time, so it’s relatively easy to capture. As for eye movements, that’s a bit more challenging because the eyes are smaller and less visible compared to the face. While eye tracking might be an interesting data source, it’s not as easy to gather compared to something like facial data or voice.''}

\subsubsection{Strengthening Biometric Data Handling and Governance to Mitigate Deepfake Threats} \label{bio_ethical_privacy}

Most EPs emphasized that improper biometric data handling, especially storing such data in cloud services, significantly increases vulnerability to Gen-AI deepfake attacks targeting personal identity. EP7 and EP9, for example, proposed a hybrid model—storing sensitive biometric data locally on-device, while allowing only encrypted backups or less-sensitive data in cloud environments. This local approach grants users \textit{``greater control''} to manage or delete their personal information, thereby reducing unauthorized access and misuse.

Many EPs then advocated robust and layered biometric handling strategies. They recommended enhancing the robustness of biometric systems through adversarial training, as it compels models to rely on inherently secure features, thus mitigating adversarial threats. For instance, EP20 highlighted the effectiveness of improved data augmentation methods to combat laundering attacks: \textit{``Incorporating diverse processing types into training data, like simulations of complex processes that can highly improves the robustness of forensic tools against unseen manipulations.''} A few EPs also proposed embedding harder-to-synthesize biometric signals, such as gaze trajectories or facial micro-expression sequences, into authentication procedures. Further, some EPs noted that high-friction authentication methods (e.g., hardware tokens, human callbacks) should be reserved exclusively for anomalous or high-risk logins rather than routine authentications; cryptographic hashes of genuine enrollment data could be securely bound to hardware roots of trust, providing tamper-evident verification. EP19 particular recommended pairing lightweight CNN filters at the edge for rapid triage with deeper, adversarial-trained detectors in cloud infrastructures, facilitating immediate responses and comprehensive forensic analysis. Furthermore, several EPs emphasized transparency and privacy in biometric data practices. EP16 proposed layered disclosures to communicate clearly about biometric data processing, enabling users to better understand potential implications related to deepfake risks. Similarly, EP8 suggested employing privacy-preserving technologies, such as differential privacy and federated learning, to minimize exposure and vulnerability of biometric data to sophisticated deepfake threats.

\textbf{Potential of continuous authentication.} 
Some EPs acknowledged the promising nature of continuous authentication in defending against Gen-AI deepfake images and videos. Instead of a single point of entry, continuous authentication relies on ongoing behavioral or physiological cues that are analyzed over the duration of a session. Like EP3 stressed, \textit{``Continuous authentication provide a better way knows who you are, by monitoring the unique behaviors of a user in real time, systems can detect anomalies that might indicate that an imposter has taken over an active session.''} However, EP24 argued that session takeover becomes feasible when continuous checks degrade to sporadic image captures rather than frame-by-frame analysis, giving attackers a window of a few seconds to replay a forged clip.

\textbf{Enhancing transparency and user consent.}
Most EPs underscored the need for transparent communication regarding the risks posed by Gen-AI-generated deepfake images and videos when collecting and storing biometric data. For instance, For instance, EP12 emphasized the necessity of transparency in preventing the misuse of deepfakes. As Gen-AI identity deepfake threats are increasingly sophisticated and difficult for users to anticipate, EP4 argued for simplifying consent mechanisms: \textit{``Information should be presented in a way that is accessible to all users, like clear, understandable interfaces rather than complex legal jargon, regardless of their technical background.''} Similarly, EP7 suggested: \textit{``standardized consent mechanisms that allow users to opt in or out of biometric authentication systems with full awareness of potential risks.''} They all noted that effective consent mechanisms should be active and unambiguous. EP7 and EP12 recommended standardized and active consent processes that users regularly revisit (e.g., utilizing privacy-by-design principles~\cite{cavoukian2021privacy}), keeping pace with evolving deepfake risks, as EP12 noting: \textit{``These interfaces should be designed to be revisited over time, allowing users to update or withdraw their consent as their circumstances or preferences change.''}

\textbf{Accountability to mitigate identity deepfake risks.} Most EPs noted the importance of accountability in biometric data management to reduce risks from Gen-AI identity deepfake threats. EP9 suggested this accountability, in turn, drives organizations to adopt more rigorous security measures and ethical data handling practices. To build an accountability framework, however, EP9 also noted that the regulatory frameworks (e.g., the GDPR, the EU AI Act) need mandate that data controllers provide clear and accessible information about data processing activities, noting: \textit{``Complying with these legal requirements should be seen as a baseline rather than the ultimate goal. This means service providers should strive to exceed regulatory minimums by embedding transparency and user consent into the core design of their biometric systems.''}

\textbf{Data governance to mitigate deepfake exploitation in biometric systems.} Notably, a few EPs recognized deepfakes as a national risk. For instance, EP21 highlighted that the DHS has warned synthetic media can exponentially challenge critical infrastructure owners and operators\footnote{\url{https://www.dhs.gov/sites/default/files/publications/increasing_threats_of_deepfake_identities_0.pdf}}, explicitly listing the use of deepfakes among AI-enabled cyberattack vectors and citing \textit{``social engineering with deepfake phishing''} as a specific threat. To address such deepfake security risks, some EPs emphasized the necessity of international cooperation, including technical requirements for data practices (e.g., encryption, storage, and transmission), protocols for auditing and compliance, and the establishment of data governance frameworks resilient to deepfake spoofing on biometric data. However, some EPs argued that forming an international consortium to harmonize standards across borders may be challenging, as each country has its own regulatory framework. For instance, EP9 highlighted the importance of developing standardized biometric security guidelines to ensure consistent safety protocols across industries, from banking to healthcare, stating: \textit{``Without a common framework, companies may implement ad hoc solutions that, while functional in isolation, create gaps when systems interact or when attackers exploit the weakest link in a multi-organizational ecosystem.''}

\textbf{Increasing public awareness and education about identity deepfake threats in biometric systems.} Most EPs acknowledged the importance of public education regarding the threats posed by Gen-AI deepfakes and its risks to biometric authentication systems, as most users remain unaware of how biometric data is captured, processed, and stored, and they may not fully appreciate the potential consequences of a breach or misuse (e.g., identity theft and fraud). Furthermore, EP7 proposed \textit{``public awareness campaigns specifically designed''} to educate users on secure biometric practices and the risks associated with deepfake manipulation through social media and public announcements. EP14 and EP15 also recommended embedding deepfake education into school curriculums to foster early awareness of privacy and security in an increasingly AI-driven world.

\textbf{Biometric authentication solution in smart home devices.} A few EPs highlighted the need for secure biometric authentication in smart home devices. However, they proposed concerns about \textit{``over collection and limit capability in data handling practices''}, especially for these device service providers. For instance, EP12 discussed the potential privacy concerns raised for users: \textit{``If a company isn’t transparent about how it collects and processes users' biometric authentication data, or even more sensitive information about their families or visitors, there’s a big risk. potentially allowing attackers to create convincing deepfakes from compromised biometric data stored or transmitted insecurely.''} Further, EP4 discussed user acceptance, noting that while users may appreciate the convenience of smart door locks with biometric authentication, many remain wary of \textit{``living under constant surveillance''} in their own homes. 

\subsubsection{Public's Views on Biometric Authentication Security and Identity Deepfake Threats} \label{public_view}

All PPs reported frequent use of fingerprint and face recognition to unlock smartphones, access banking apps, and for identity checks (e.g., at airports or customs). They noted that biometric authentication is significantly more convenient than passwords, particularly for mobile banking and device access. Face ID was often favored for its hands-free nature. PP3 and PP7 mentioned having used iris authentication in government offices, and PP6 noted experience with vein authentication in a smart door lock. Others were unfamiliar with advanced biometric modalities, such as gait and iris biometrics.

Despite nearly half of PPs recognizing that hardware-based authentication (e.g., USB keys) could potentially offer strong security, like PP1 noted: \textit{``I think convenience and security are opposites. The more troublesome the biometric technology is, the safer it is.''} While they prioritized convenience over privacy and security, stating that biometric authentication eliminates the need for remembering passwords. For instance, PP6 indicated they would revert to passwords if biometric security were compromised.

Most PPs viewed deepfake images/videos as an entertainment way. However, PP4 highlighted a more serious concern about fueling conspiracy theories by referencing the AI-generated image of Catherine, Princess of Wales, during her public absence, emphasizing the ethical issues AI-generated content may pose: \textit{``I think one of the key things is that we have a picture or video to prove something it actually happened, or something it's true. I can't imagine one day that generated video becomes unidentified from the real one and what will happen to, a big trouble to laws and criminal investigation. Then what can actually be used as an evidence?''} Similarly, PP12 expressed concerns about being targeted by Gen-AI deepfake images and videos, particularly deepfaked adult and explicit contents: \textit{``Your face data is way too easy to collect now. I never thought I’d have to worry about someone using my photos on Instagram to fake my face and make porn. But it’s happening, and it’s honestly terrifying. How are we supposed to be protected when the laws haven’t even caught up yet?''}

Although most PPs lacked a technical understanding of Gen-AI deepfake spoofing on personal identity, several worried that deepfake could potentially replicate biometric features, making face authentication less reliable (e.g., PP4, PP7). More than half PPs noted that facial authentication might be susceptible to hacking, often referencing scam or fraud stories they had come across. In contrast, and mirroring the views of the experts, most public participants perceived iris recognition as the strongest safeguard, citing its uniqueness and the perceived difficulty of replication.Some PPs assumed that if banks, reputable companies, or government services endorsed biometrics, then it must be secure, exhibiting an externalized trust model in which users rely on institutional credibility more than their own security vigilance~\cite{mcknight2002developing}.  However, a few participants were more skeptical about lesser-known companies’ capabilities to securely store biometric data, highlighting the biometric data storage depending on contexts. Some PPs also expressed worries about government misuse or insufficient regulations, fearing data leaks or unauthorized surveillance. As PP8 stated: \textit{``Privacy breaches are just part of life, if the government wants your data, they’ll get it, with or without your permission. But what can we do? Most of us don’t really take any steps to protect ourselves anyway.''}

\section{Discussion}
\label{discussion}

\subsection{Summary of Findings}
Our EPs acknowledged biometric authentication as a valuable safeguard, yet strongly cautioned about the rapidly evolving threats from advanced Gen-AI deepfake technologies. They highlighted the reduced barriers to deepfake generation, noting that attackers can now reliably create convincing synthetic identities using minimal publicly available media and inexpensive cloud resources, significantly amplifying risks across financial, reputational, and critical infrastructure sectors. They identified substantial limitations in current biometric security practices, especially traditional modalities like facial and voice recognition, emphasizing their vulnerability to spoofing from sophisticated deepfake manipulations capable of mimicking subtle physiological signals such as pupil dilation and microsaccades. EPs further noted that prevailing detection methods, predominantly deep learning-based, often lack robustness and generalizability against realistic adversarial conditions, becoming ineffective when encountering altered resolutions, compressions, or media laundering attacks. Consequently, experts recommended adopting dynamic, involuntary biometric signals (e.g., eye movements or facial micro-expressions) as these are inherently more challenging to replicate. They also advocated multilayered defensive approaches, including adversarial training, diverse data augmentation, privacy-preserving technologies, continuous authentication mechanisms, clear user consent procedures, rigorous data governance frameworks, international cooperation, and proactive public education to significantly enhance preparedness and resilience against emerging deepfake-driven biometric threats. \textbf{(RQ1)}

Our PPs generally trusted biometric authentication (frequently citing convenience for smartphone/smart home device unlocking or banking apps), only a minority understood how deepfakes might undermine facial or voice recognition. We found that younger, recently educated, and less experienced professionals (especially from Academia and Tech) have significantly greater AI familiarity, indicating generational rather than experiential effects. Academia maintains notably more critical attitudes toward F4 and F5 compared to other industries; healthcare uniquely reports higher perceptions in F2. However, bachelor's degree holders exhibit higher practical AI familiarity (F1) than advanced-degree holders, suggesting specialized higher education may not necessarily enhance practical AI knowledge. We also found strong interconnections between factors (e.g., F1 and F2). Mediation effects highlight familiarity (F1) as a pivotal mediator influencing F4 and F5. Across all demographic groups, misconceptions persist about how easily deepfakes can manipulate personal biometric data, underscoring experts’ emphasis on broader public education. \textbf{(RQ2)} 

\subsection{Contribution to Prior Work}
Our findings extend previous research on biometric security and deepfake threats by systematically synthesizing insights from expert assessments and public perceptions, directly addressing RQ1 and RQ2. While earlier studies primarily focused either on biometric trust issues (e.g.,~\cite{alrawili2024comprehensive,rui2018survey}) and deepfake misinformation risks (e.g.,~\cite{chesney2019deep,gregory2022deepfakes}), our research uniquely explores their intersection through empirical data from both experts and the public.

\textbf{Empirical evidence of perceptions and vulnerabilities.} 
Our survey and EP interviews reveal nuanced perceptions regarding biometric authentication and deepfake threats. Our study significantly enhances prior biometric threat models by explicitly incorporating expert-driven insights into a structured deepfake kill-chain~\cite{hutchins2011intelligence}. While traditional cybersecurity frameworks such as the cyber-kill-chain focus broadly on cyberattacks, our work uniquely adapts and expands this model to systematically detail how Gen-AI deepfake technologies specifically threaten biometric authentication. This enriched threat model offers actionable insights into the vulnerabilities at each attack stage, along with empirically grounded recommendations for mitigating these threats through dynamic biometrics, layered detection approaches, and improved governance practices (see \S \ref{deepfake_threat_model}). 

Although Srinivasan's work reported no significant demographic effects on privacy concern, although awareness correlated positively with comfort using biometrics~\cite{srinivasan2023understanding}. By contrast, our larger, cross-sector survey sample uncovers pronounced age, industry and education gradients in AI familiarity, ethical concern and confidence, showing that demographic factors can powerfully shape perceptions of deepfake-related biometric risk. Further, Kaate et al. explored users’ reactions to deepfake personas, surfacing qualitative themes of realism, trust, and distracting artifacts~\cite{kaate2023users}. Our quantitative findings corroborate their observations of trust erosion when synthetic cues are detected, yet we extend the analysis to the biometric authentication context, quantify these perceptions across sectors, and link them to actionable mitigation preferences (see \S \ref{mitigations_for_threats}). 

\textbf{Proposing dynamic biometrics against identity deepfakes.}
Building upon existing studies advocating for dynamic and behavioral biometrics (e.g., facial micro-expressions~\cite{saeed2021facial}, eye blinking~\cite{jung2020deepvision, li2018ictu}, eye movements~\cite{lei2023end}, gesture~\cite{simske2009dynamic}), our results empirically reinforce the potential of these involuntary signals as more resilient against deepfake threats~\cite{pei2024deepfake, shahzad2022review, 9010912}. Experts explicitly identified facial micro-expression sequences and eye-movement trajectories as particularly resistant to current Gen-AI spoofing attempts due to their complex, user-specific, and difficult-to-replicate nature~\cite{jung2020deepvision, haliassos2021lips}. However, our findings uniquely combine feasibility assessments with usability considerations, highlighting practical limitations in consumer adoption, such as authentication time constraints and difficulties in data collection, previously unaddressed in technical literature. This study provides a practical roadmap for integrating advanced biometric modalities into mainstream authentication systems.

\textbf{Framework for biometric data governance and education strategies.}
Our research also identifies critical gaps in biometric data handling practices and offers a novel framework that prioritizes local storage, robust adversarial training, and comprehensive public education. Experts recommended hybrid storage models (combining on-device sensitive data storage with encrypted cloud backups), along with layered security measures (e.g., adversarial training, privacy-preserving technologies such as differential privacy and federated learning) to enhance robustness against sophisticated deepfake manipulations (see \S \ref{bio_ethical_privacy}). Crucially, both EPs and PPs emphasized a strong need for clearer transparency mechanisms and user-friendly consent frameworks to ensure users fully understand biometric data handling implications and associated risks. Furthermore, respondents across sectors highlighted the urgent requirement for targeted public education and awareness campaigns—an aspect largely underexplored in prior biometric authentication literature—to bridge knowledge gaps and promote user preparedness against evolving Gen-AI deepfake threats.

\subsection{Deepfake Threat Modeling and Mitigation Framework to Biometric Authentication} \label{deepfake_threat_model}

\subsubsection{Deepfake Kill Chain Threat Model}
Our findings extend existing threat-modeling frameworks by integrating insights from our EPs regarding the intersection between deepfake technologies and biometric authentication systems. Drawing on the intrusion kill chain~\cite{hutchins2011intelligence}, our results (see \S \ref{deepfake_threats}) highlight critical vulnerabilities across multiple phases of biometric authentication attacks. Rapid advancements in Gen-AI significantly simplify the creation and deployment of highly realistic deepfakes, dramatically reducing technical barriers for adversaries, particularly during the \textit{Weaponization}, \textit{Delivery}, \textit{Exploitation}, and \textit{Installation} phases. For instance, EPs mentioned that attackers leverage increasingly streamlined profiling methods by harvesting publicly available multimedia sources, enabling the rapid production of convincing identity clones within mere hours via inexpensive cloud services. Furthermore, sophisticated synthetic asset generation techniques—such as NeRF and diffusion models—pose severe threats to the reliability of traditional biometric measures, notably static facial and voice authentications. Such advancements amplify risks from high-value national targets, including critical infrastructures and key personnel, down to general public users, making initial \textit{Reconnaissance} activities significantly more accessible and cost-effective. Our empirical data also indicates that public perceptions significantly lag behind expert assessments, especially regarding the understanding of deepfake-related risks. Public users prioritize convenience over security, trust institutional endorsements rather than practicing personal vigilance, and generally remain unaware of the severity of deepfake threats (see \S \ref{public_view}). This public gap in awareness facilitates easier adversary operations in the \textit{Command and Control (C2)} phase by exploiting users' limited readiness and insufficient knowledge. 

\subsubsection{Mitigation Framework}\label{mitigations_for_threats}
Our mitigation framework addresses vulnerabilities across the deepfake kill chain threat model from technical, social, and legal perspectives.

\textbf{Technical mitigation.} Targeting phases of \textit{Weaponization}, \textit{Delivery}, \textit{Exploitation}, and \textit{Installation}, our findings emphasize transitioning away from static biometric methods towards dynamic, behavior-based modalities. Embedding dynamic biometric signals, such as involuntary eye movements (e.g., micro-saccade and drift), offers a promising defense due to the inherent difficulty current Gen-AI models face replicating these continuous, nuanced data patterns~\cite{lei2023end}. Editable biometric traits like gaze trajectories and device interaction patterns further mitigate replication risks by allowing periodic updates. Implementing layered defenses, such as MFA, anomaly detection, and real-time deepfake detection at both edge and cloud infrastructure levels, is critical for robust protection against spoofing and manipulation attacks. Transparent disclosure of biometric data handling practices, along with regular user notifications about deepfake risks, is essential, especially for high-stakes environments such as banking transactions, governmental services, and virtual meetings (e.g., Zoom).

\textbf{Social mitigation.} Addressing vulnerabilities from \textit{Reconnaissance} through \textit{Command and Control (C2)}, our qualitative results underscore the persistent gaps in user understanding and consent regarding biometric data use. We advocate clear, concise consent interfaces enabling users to manage biometric permissions proactively, particularly in multi-user contexts such as smart homes~\cite{abdi2021privacy}. Targeted public education initiatives should bridge demographic divides, focusing especially on older or experienced professionals in sectors such as government and healthcare, thereby improving readiness against deepfake threats. Awareness campaigns leveraging accessible, engaging formats, infographics, videos, and social media, can effectively communicate risks and appropriate mitigation measures. Integrating digital literacy into education curricula and mandatory professional training will further enhance the public's ability to recognize and respond proactively to deepfake threats. Civil society and non-governmental organizations (NGOs) should partner to provide neutral, practical resources and community-driven response plans to bolster societal resilience.

\textbf{Legal and regulatory mitigation.} To fortify defenses against vulnerabilities identified across the kill chain phases, including \textit{Weaponization}, \textit{Delivery}, and \textit{Exploitation}, robust legal and regulatory frameworks must be established. Policymakers and organizations should adopt stringent privacy-preserving biometric practices consistent with regulations like the GDPR and the EU AI Act. Enhanced accountability through substantial penalties for inadequate biometric data protection and unauthorized use is critical. International cooperation, through global treaties and certifications, should standardize biometric data handling rules, facilitating cross-border enforcement and compliance~\cite{chin2022governing, mitchell2017don,xu2024global, romero2024generative}. Furthermore, addressing gender-specific abuses (e.g., non-consensual explicit deepfakes) through strengthened legal frameworks requires ongoing research, particularly involving victim perspectives to ensure rights and dignity are effectively safeguarded~\cite{sippy2024behind,akter2025emergence}.

\subsection{Limitations} 
Our study has several limitations. First, our EP interviews may have biased discussions toward more technical perspectives. While a few addressed AI regulations, future research should incorporate experts from other disciplines (e.g., sociology and law) and junior researchers that could provide more practical insights. Also, our PPs were mainly recruited from university, limiting demographic and cultural diversity (e.g., age and education levels). However, our survey includes a broader demographic range that helps fill this gap. As our study of the PPs is still UK-centric, future research might systematically explore differences between countries, assessing how regional variations in regulation, technology adoption, and cultural norms influence perceptions and practices. Secondly, self-selection bias may have influenced our findings, as individuals with a strong interest in AI \& biometric were more likely to volunteer, potentially skewing the range of perspectives~\cite{bethlehem2010selection}. In addition, self-reported data inherently carries the risk of social desirability bias, where participants may have provided responses they believed were more acceptable rather than their genuine views~\cite{nederhof1985methods}. Lastly, our exclusive focus on image- and video-based identity deepfakes; while other emerging deepfake modalities (e.g., text, audio, hypermedia) remain outside the scope of this study. Future research can encompass a broader spectrum of deepfakes and investigate their associated security and privacy implications for the public. 

\subsection{Future Work}

\textbf{Advancing dynamic \& editable biometric modalities.} Based on \S \ref{deepfake_threat_model} and \S \ref{findings}, future research should prioritize empirical validation of dynamic behavioral biometric modalities, especially involuntary eye movements and facial micro-expressions, in realistic consumer device scenarios. Given the identified vulnerabilities, future research should move beyond single-point, static biometric verification methods due to rapidly advancing Gen-AI deepfake technologies: dynamic, behavior-based, and editable biometric modalities (e.g., eye movements) should be prioritized as complementary authentication methods to traditional biometrics, particularly on widely used consumer devices (e.g., smartphones)~\cite{lei2023end,lei2023dynamicread,katsini2020role}. Building upon lab-based results (e.g.,~\cite{ivo2018fastreplay}), further studies should aim for viable accuracy and practical application. Additionally, targeted interventions to address demographic disparities in AI familiarity and preparedness, particularly among mid-to-late career professionals, warrant exploration. Cross-sector comparative analyses could uncover barriers limiting AI adoption and facilitate targeted educational and policy initiatives.

\textbf{Contextual authentication in smart homes.}
According to \S \ref{bio_ethical_privacy}, we found that risks associated with biometric data in multi-user smart homes are vulnerable to Gen-AI deepfake compromises~\cite{lee2024deepfakes,bilika2023hello}. Given the complexities in multi-user scenarios, varying security thresholds, privacy expectations, and potential biometric spoofing, future studies should design user-centric, context-aware biometric frameworks specifically tailored to diverse household environments. Addressing vulnerabilities such as compromised biometric data and sophisticated spoofing attacks should be integral~\cite{zimmermann2024authenticate,thakkar2022would,pattnaik2024security,zhang2022teo}. 

\textbf{Public awareness and educational initiatives}
Given the public knowledge gap highlighted in \S \ref{survey_results} and \S \ref{public_view}, extensive educational initiatives about Gen-AI deepfake risks are urgently needed, particularly within high-security sectors (e.g., banking, e-government, healthcare, and tech industries)~\cite{uddagiri2024ethical,gregory2023fortify,apolo2024beyond}. Awareness programs should initially target specific scenarios vulnerable to social engineering and impersonation attacks (e.g., virtual meetings, remote authentications, online service authorizations) before broadly expanding to the general population. Future research should also systematically assess these programs' effectiveness, creating tailored strategies to mitigate misconceptions and foster proactive security behaviors against deepfake threats. Future research should explore strengthening existing legal frameworks to better protect individuals, particularly women, from deepfake-related abuses.

\section{Conclusion}
\label{conclusion}

This study provides empirical evidence to show how both experts and the general public view the intersection of Gen-AI identity deepfake threats
and biometric authentication systems. Firstly, our EPs acknowledged biometric authentication's utility yet cautioned against the swiftly evolving threats posed by advanced Gen-AI deepfake technologies. They emphasized reduced barriers for deepfake generation, noting adversaries now create highly realistic synthetic identities quickly using publicly available media and inexpensive cloud resources, amplifying risks across critical sectors. EPs highlighted vulnerabilities in traditional biometric methods, particularly facial and voice recognition, underscoring the limitations of current detection methods in realistic adversarial conditions. Consequently, experts recommended dynamic, editable biometric signals such as eye movements and multilayered defensive strategies, including continuous authentication, transparent consent procedures, robust governance frameworks, international cooperation, and proactive public education, to bolster resilience. 

Secondly, the public interviews shows moderate trust in biometrics, driven primarily by convenience rather than a deep understanding of the threat landscape. Our quantitative results reveal significant demographic stratification and complex interactions shaping perceptions of AI familiarity, trust in biometrics, industry readiness, biometric security confidence, and ethical AI adoption. In particular, public responses generally trusted biometric authentication for convenience but exhibited limited understanding of deepfake risks, indicating generational differences in AI familiarity. Younger, recently educated professionals demonstrated higher practical AI familiarity, particularly in academia and technology sectors. Academia notably exhibited critical attitudes towards biometric security practices, emphasizing ethical considerations. Across demographic groups, persistent misconceptions about deepfake risks highlight a crucial need for comprehensive public education.

Lastly, we extend existing threat models by integrating expert-informed insights into a structured deepfake kill-chain, to systematically illustrate how Gen-AI deepfake technologies pose targeted threats to biometric authentication systems. To address these deepfake kill-chain threats, we propose a mitigation framework from technical, legal, and social objectives, calling for a more comprehensive defense approach that integrates human, technical, and policy dimensions. By fostering multi-factor, context-aware authentication strategies and equipping users with greater awareness, stakeholders can work collectively to mitigate Gen-AI deepfake abuses while preserving the usability and accessibility that define more secure biometric solutions.

\bibliographystyle{unsrtnat}
\bibliography{references}  






\appendix
\section{Survey Data Analysis}
\subsection{Biometric Authentication Usage}

\subsection{Demographic Differences}
\begin{table}[ht]
\centering
\fontsize{6pt}{6pt}\selectfont
\resizebox{0.48\textwidth}{!}{%
\begin{tabular}{llcccc}
\toprule
\textbf{Group 1} & \textbf{Group 2} & \textbf{Mean Diff.} & \textbf{Adj. p-value} & \textbf{95\% CI Lower} & \textbf{95\% CI Upper}\\
\midrule
18-24 & 25-34 & -0.081 & 0.960 & -0.405 & 0.243\\
18-24 & 35-44 & -0.589 & *** & -0.933 & -0.245\\
18-24 & 45-54 & -0.778 & *** & -1.161 & -0.395\\
18-24 & 55+ & -1.136 & *** & -1.548 & -0.723\\
25-34 & 35-44 & -0.508 & *** & -0.728 & -0.288\\
25-34 & 45-54 & -0.697 & *** & -0.974 & -0.420\\
25-34 & 55+ & -1.055 & *** & -1.371 & -0.738\\
35-44 & 45-54 & -0.189 & 0.418 & -0.490 & 0.111\\
35-44 & 55+ & -0.547 & *** & -0.884 & -0.210\\
45-54 & 55+ & -0.358 & 0.072 & -0.735 & 0.019\\
\bottomrule
\multicolumn{6}{l}{\textit{Note:} Significance codes: * p $<$ 0.05, ** p $<$ 0.01, *** p $<$ 0.001}
\end{tabular}}
\caption{Tukey's HSD Post-Hoc Analysis for F1\_AI\_Familiarity by Age}
\label{tab:posthoc_f1_age}
\end{table}

\begin{table}[ht]
\centering
\fontsize{6pt}{6pt}\selectfont
\resizebox{0.48\textwidth}{!}{%
\begin{tabular}{llcccc}
\toprule
\textbf{Group 1} & \textbf{Group 2} & \textbf{Mean Diff.} & \textbf{Adj. p-value} & \textbf{95\% CI Lower} & \textbf{95\% CI Upper}\\
\midrule
Academia & Finance & -0.432 & *** & -0.664 & -0.200\\
Academia & Gov./Public & -0.302 & ** & -0.514 & -0.089\\
Academia & Healthcare & -0.455 & *** & -0.709 & -0.201\\
Academia & Tech & -0.304 & ** & -0.523 & -0.086\\
Finance & Gov./Public & 0.130 & 0.306 & -0.055 & 0.315\\
Finance & Healthcare & -0.023 & 0.999 & -0.255 & 0.209\\
Finance & Tech & 0.127 & 0.365 & -0.065 & 0.319\\
Gov./Public & Healthcare & -0.153 & 0.277 & -0.366 & 0.059\\
Gov./Public & Tech & -0.003 & 1.000 & -0.170 & 0.164\\
Healthcare & Tech & 0.151 & 0.323 & -0.067 & 0.369\\
\bottomrule
\multicolumn{6}{l}{\textit{Note:} Significance codes: * p $<$ 0.05, ** p $<$ 0.01, *** p $<$ 0.001}
\end{tabular}}
\caption{Tukey's HSD Post-Hoc Analysis for F5\_Ethical\_AI\_Adoption by Working Area}
\label{tab:posthoc_f5_workingarea}
\end{table}

\begin{table}[ht]
\centering
\fontsize{6pt}{6pt}\selectfont
\resizebox{0.48\textwidth}{!}{%
\begin{tabular}{llcccc}
\toprule
\textbf{Group 1} & \textbf{Group 2} & \textbf{Mean Diff.} & \textbf{Adj. p-value} & \textbf{95\% CI Lower} & \textbf{95\% CI Upper}\\
\midrule
Bachelor & Doctoral & 0.433 & * & 0.072 & 0.793\\
Bachelor & $\leq$ High School & -0.528 & * & -0.977 & -0.078\\
Bachelor & Master & 0.331 & *** & 0.109 & 0.553\\
Doctoral & $\leq$ High School & -0.960 & *** & -1.514 & -0.406\\
Doctoral & Master & -0.102 & 0.909 & -0.494 & 0.291\\
$\leq$ High School & Master & 0.858 & *** & 0.383 & 1.334\\
\bottomrule
\multicolumn{6}{l}{\textit{Note:} Significance codes: * p $<$ 0.05, ** p $<$ 0.01, *** p $<$ 0.001}
\end{tabular}}
\caption{Tukey's HSD Post-Hoc Analysis for F1\_AI\_Familiarity by Education Degree}
\label{tab:posthoc_f1_education}
\end{table}

\begin{table*}[ht]
\centering
\fontsize{6pt}{6pt}\selectfont
\resizebox{0.5\textwidth}{!}{%
\begin{tabular}{llrrrrr}
\toprule
\textbf{Predictor} & \textbf{Outcome} & \textbf{Coefficient} & \textbf{p-value} & \textbf{R²} & \textbf{F-statistic} & \textbf{F p-value} \\
\midrule
\multirow{3}{*}{F1} & F2 & 0.243 & $<$ 0.001*** & 0.150 & 71.82 & $<$ 0.001*** \\
 & F4 & 0.289 & $<$ 0.001*** & 0.142 & 67.13 & $<$ 0.001*** \\
 & F5 & 0.208 & $<$ 0.001*** & 0.104 & 47.06 & $<$ 0.001*** \\
\midrule
\multirow{4}{*}{F2} & F1 & 0.619 & $<$ 0.001*** & 0.150 & 71.82 & $<$ 0.001*** \\
 & F3 & 0.353 & $<$ 0.001*** & 0.058 & 25.20 & $<$ 0.001*** \\
 & F4 & 0.427 & $<$ 0.001*** & 0.122 & 56.57 & $<$ 0.001*** \\
 & F5 & 0.321 & $<$ 0.001*** & 0.097 & 43.42 & $<$ 0.001*** \\
\midrule
\multirow{3}{*}{F3} & F2 & 0.165 & $<$ 0.001*** & 0.058 & 25.20 & $<$ 0.001*** \\
 & F4 & 0.094 & 0.023* & 0.013 & 5.18 & 0.023* \\
 & F5 & 0.117 & $<$ 0.001*** & 0.027 & 11.45 & $<$ 0.001*** \\
\midrule
\multirow{4}{*}{F4} & F1 & 0.492 & $<$ 0.001*** & 0.142 & 67.13 & $<$ 0.001*** \\
 & F2 & 0.286 & $<$ 0.001*** & 0.122 & 56.57 & $<$ 0.001*** \\
 & F3 & 0.134 & 0.023* & 0.013 & 5.18 & 0.023* \\
 & F5 & 0.210 & $<$ 0.001*** & 0.062 & 26.75 & $<$ 0.001*** \\
\midrule
\multirow{4}{*}{F5} & F1 & 0.499 & $<$ 0.001*** & 0.104 & 47.06 & $<$ 0.001*** \\
 & F2 & 0.301 & $<$ 0.001*** & 0.097 & 43.42 & $<$ 0.001*** \\
 & F3 & 0.235 & $<$ 0.001*** & 0.027 & 11.45 & $<$ 0.001*** \\
 & F4 & 0.295 & $<$ 0.001*** & 0.062 & 26.75 & $<$ 0.001*** \\
\bottomrule
\multicolumn{7}{l}{\textit{Note:} Significance codes: * p $<$ 0.05, ** p $<$ 0.01, *** p $<$ 0.001} \\
\end{tabular}}
\caption{Simple Linear Regression Results for Relationships Among AI and Biometric Factors}
\label{tab:regression_results}
\end{table*}

\begin{table*}[!htbp]
\centering
\fontsize{6pt}{6pt}\selectfont
\resizebox{0.5\textwidth}{!}{%
\begin{tabular}{lllrrr}
\toprule
\textbf{Mediation Path} & \textbf{Path Component} & \textbf{Coefficient} & \textbf{p-value} & \textbf{R²} & \textbf{Significance} \\
\midrule
\multirow{3}{*}{F2 $\rightarrow$ F1 $\rightarrow$ F4} & Path a (F2 $\rightarrow$ F1) & 0.619 & $<$ 0.001 & 0.150 & *** \\
 & Path b (F1 $\rightarrow$ F4) & 0.289 & $<$ 0.001 & 0.142 & *** \\
 & Direct path (F2 $\rightarrow$ F4) & 0.427 & $<$ 0.001 & 0.122 & *** \\
\midrule
\multirow{3}{*}{F3 $\rightarrow$ F2 $\rightarrow$ F4} & Path a (F3 $\rightarrow$ F2) & 0.165 & $<$ 0.001 & 0.058 & *** \\
 & Path b (F2 $\rightarrow$ F4) & 0.427 & $<$ 0.001 & 0.122 & *** \\
 & Direct path (F3 $\rightarrow$ F4) & 0.094 & 0.023 & 0.013 & * \\
\midrule
\multirow{3}{*}{F1 $\rightarrow$ F5 $\rightarrow$ F3} & Path a (F1 $\rightarrow$ F5) & 0.208 & $<$ 0.001 & 0.104 & *** \\
 & Path b (F5 $\rightarrow$ F3) & 0.235 & $<$ 0.001 & 0.027 & *** \\
 & Direct path (F1 $\rightarrow$ F3) & -- & -- & -- & -- \\
\midrule
\multirow{3}{*}{F4 $\rightarrow$ F2 $\rightarrow$ F5} & Path a (F4 $\rightarrow$ F2) & 0.286 & $<$ 0.001 & 0.122 & *** \\
 & Path b (F2 $\rightarrow$ F5) & 0.321 & $<$ 0.001 & 0.097 & *** \\
 & Direct path (F4 $\rightarrow$ F5) & 0.210 & $<$ 0.001 & 0.062 & *** \\
\midrule
\multirow{3}{*}{F5 $\rightarrow$ F4 $\rightarrow$ F1} & Path a (F5 $\rightarrow$ F4) & 0.295 & $<$ 0.001 & 0.062 & *** \\
 & Path b (F4 $\rightarrow$ F1) & 0.492 & $<$ 0.001 & 0.142 & *** \\
 & Direct path (F5 $\rightarrow$ F1) & 0.499 & $<$ 0.001 & 0.104 & *** \\
\bottomrule
\multicolumn{6}{l}{\textit{Note:} Significance codes: * p $<$ 0.05, ** p $<$ 0.01, *** p $<$ 0.001} \\
\end{tabular}}
\caption{Selected Significant Mediation Pathways Among AI and Biometric Factors}
\label{tab:mediation_results}
\end{table*}

\begin{table*}[ht]
\centering
\fontsize{6pt}{6pt}\selectfont
\resizebox{0.5\textwidth}{!}{%
\begin{tabular}{llrrrr}
\toprule
\textbf{Interaction Model} & \textbf{Predictor} & \textbf{Coefficient} & \textbf{p-value} & \textbf{R²} & \textbf{F-statistic} \\
\midrule
\multirow{4}{*}{F3 $\times$ F5 $\rightarrow$ F2} & Constant & 0.143 & 0.839 & \multirow{4}{*}{0.144} & \multirow{4}{*}{22.63} \\
 & F3 & 0.631 & 0.007** & & \\
 & F5 & 0.714 & $<$ 0.001*** & & \\
 & F3 $\times$ F5 Interaction & -0.147 & 0.030* & & \\
\midrule
\multirow{4}{*}{F3 $\times$ F2 $\rightarrow$ F5} & Constant & 0.837 & 0.180 & \multirow{4}{*}{0.118} & \multirow{4}{*}{18.09} \\
 & F3 & 0.552 & 0.006** & & \\
 & F2 & 0.805 & $<$ 0.001*** & & \\
 & F3 $\times$ F2 Interaction & -0.166 & 0.015* & & \\
\midrule
\multirow{4}{*}{F3 $\times$ F1 $\rightarrow$ F4} & Constant & 1.429 & $<$ 0.001*** & \multirow{4}{*}{0.169} & \multirow{4}{*}{27.37} \\
 & F3 & 0.359 & 0.002** & & \\
 & F1 & 0.656 & $<$ 0.001*** & & \\
 & F3 $\times$ F1 Interaction & -0.118 & 0.025* & & \\
\midrule
\multirow{4}{*}{F5 $\times$ F2 $\rightarrow$ F1} & Constant & 5.183 & $<$ 0.001*** & \multirow{4}{*}{0.226} & \multirow{4}{*}{39.30} \\
 & F5 & -1.326 & 0.002** & & \\
 & F2 & -1.386 & 0.004** & & \\
 & F5 $\times$ F2 Interaction & 0.560 & $<$ 0.001*** & & \\
\midrule
\multirow{4}{*}{F2 $\times$ F4 $\rightarrow$ F1} & Constant & 2.776 & 0.012* & \multirow{4}{*}{0.232} & \multirow{4}{*}{40.67} \\
 & F2 & -0.563 & 0.127 & & \\
 & F4 & -0.594 & 0.081 & & \\
 & F2 $\times$ F4 Interaction & 0.318 & 0.005** & & \\
\bottomrule
\multicolumn{6}{l}{\textit{Note:} Significance codes: * p $<$ 0.05, ** p $<$ 0.01, *** p $<$ 0.001} \\
\end{tabular}}
\caption{Significant Interaction Effects Among AI and Biometric Factors}
\label{tab:interaction_results}
\end{table*}

\end{document}